\pdfoutput=1

\documentclass{egpubl}
\usepackage{eurovis2020}
\usepackage[T1]{fontenc}
\usepackage{dfadobe}
\usepackage{cite}  % comment out for biblatex with backend=biber

\BibtexOrBiblatex
\electronicVersion
\PrintedOrElectronic
\ifpdf \usepackage[pdftex]{graphicx} \pdfcompresslevel=9
\else \usepackage[dvips]{graphicx} \fi

\usepackage{amsmath} % assumes amsmath package installed
\usepackage{amssymb}  % assumes amsmath package installed
\usepackage{textcomp}
\usepackage{flafter}
\usepackage{caption}
\usepackage{subcaption}
\usepackage{float}
\usepackage{wrapfig}
\usepackage{lscape}
\usepackage{rotating}
\usepackage{epstopdf}
\usepackage{egweblnk}
\usepackage{xcolor}
\title[Visualisation of Kinematic Data from a Motor Tele-rehabilitation System]%
      {Web Platform for Visualisation of Kinematic Data captured from a Motor Tele-rehabilitation System }

\author[Praveena Satkunarajah\& Kat Agres]
{\parbox{\textwidth}{\centering Praveena Satkunarajah
$^{1}$
        \& Kat Agres$^{2}$
        }
      \\
{\parbox{\textwidth}{\centering $^1$ Faculty of Medicine, Memorial University of Newfoundland, St. John's, NL, Canada\\
         $^2$Yong Siew Toh Conservatory of Music, National University of Singapore, Singapore
      }
}
}

\begin{document}

\maketitle
\begin{abstract}
Stroke can have a severe impact on an individual's quality of life, leading to consequences such as motor loss and communication problems, especially among the elderly. Studies have shown that early and easy access to stroke rehabilitation can improve an elderly individual's quality of life, and that telerehabilitation is a solution that facilitates this improvement. In this work, we visualize movement to music during rehabilitation exercises captured by the Kinect motion sensor, using a dedicated Serious Game called `Move to the Music'(MoMu). Our system provides a quantitative view of progress made by patients during a motor rehabilitation regime for healthcare professionals to track remotely (tele-rehab). \\

\printccsdesc   
\end{abstract}  
%-------------------------------------------------------------------------
\section{Introduction}

Stroke can have a severe impact on an individual's quality of life, leading to consequences such as motor loss and communication problems, especially among the elderly \cite{Saposnik2008}. Studies have shown that while early and easy access to stroke rehabilitation can improve an elderly individual's quality of life, many patients do not attend outpatient rehabilitation sessions due to restrictions in mobility and access to transportation \cite{Lui2018}. Tele-rehabilitation is a potential solution for these issues, because it brings rehabilitation to patients, rather than the other way around, providing flexibility in terms of time and location \cite{HungKN2019,
CHEN201911}.

An approach to motor rehabilitation that has been widely effective, particularly in the case of gait rehabilitation among patients with Parkinson's disease or multiple sclerosis, is the use of rhythmic auditory motor entrainment. Rhythmic entrainment is the process in which the frequency of two autonomous systems become synchronised through interaction \cite{trost2017rhythmic}. Thaut and colleagues have \cite{thaut1999rhytmicity} shown that the periodicity of auditory rhythmic patterns can entrain the human motor system, and improve movement patterns in patients with movement disorders by providing information to the brain for optimising movement. 
In addition to improving gait \cite{McIntosh22,Gui-binSong,ghai2018effects}, rhythmic auditory cueing has also been shown to improve performance in upper extremity movements in stroke rehabilitation. An experiment in \cite{thaut2002kinematic} showed that rhythmic entrainment reduced variability in timing and displacement of reaching trajectories. 
An immediate reduction in variability of arm kinematics was also shown, and acceleration and velocity profiles of the wrist joint showed significant smoothing. That is, rhythmic entrainment exercises seem to improve the stability and smoothness of reaching movements. Thaut and colleagues also found that motor response is synchronised to the rhythmic period during entrainment, and is maintained even in the event of slight frequency changes. In light of this finding, periodicity is one of the measures that our tool computes and visualises to quantify progress in quality.  

There have been several proposed tools on the use of visual analysis for motor rehabilitation. An analysis and visualisation system designed by Rahman visualised data such as change in angle joints over time, change in velocity over time and distance between relevant joints (i.e. shoulder and wrist for flexion actions) ~\cite{rahman2015multimedia}. The system also provides information on the number of correct events, and range of motion. Movement data was captured using the Kinect. In another system, Purwantiningsih et. al categorised movements as positive, negative, and neutral event types ~\cite{Purwantiningsih2016}. The visualisation interface displays the number of events, type of events over a particular movement area, and whether these patterns vary over movement areas and sessions. These tools do not, however, cater to the use of music in rehabilitation, or provide fine-grained insight on the quality of movement.

In this work, we visualize movement to music during rehabilitation exercises as part of a dedicated Serious Game called `Move to the Music' (MoMu) \cite{agres2017music}, a medtech game in which participant's movements are captured by the Microsoft Kinect Sensor V2. The Kinect for Windows SDK 2.0 allows for skeletal tracking for up to 25 joints per person, which we make use of in this system. Information for the Kinect V2 can be found at \httpAddr{//microsoft.com/en-us/research/video/project-kinect-for-azure-depth-sensor-technology/}.
Our tool differs from the above-mentioned systems by providing fine-grained details about users' quality of movements and progress across sessions using multiple kinematic measures. In addition, because MoMu is a system that centers around rhythmic entrainment for motor rehabilitation, the employed metrics quantify consistency and smoothness of the user's actions.

This tool is able to provide a quantitative view of progress of movement quality made by patients undergoing motor rehabilitation for healthcare professionals to track remotely (e.g, via the web for tele-rehabilitation).  We believe this platform provides a novel way of visualising music-entrained kinematic data, providing valuable feedback to clinicians, and supporting patients in their rehabilitation journey.

\section{SYSTEM OVERVIEW}

\begin{figure}[htp]
    \centering
    \includegraphics[width=\linewidth]{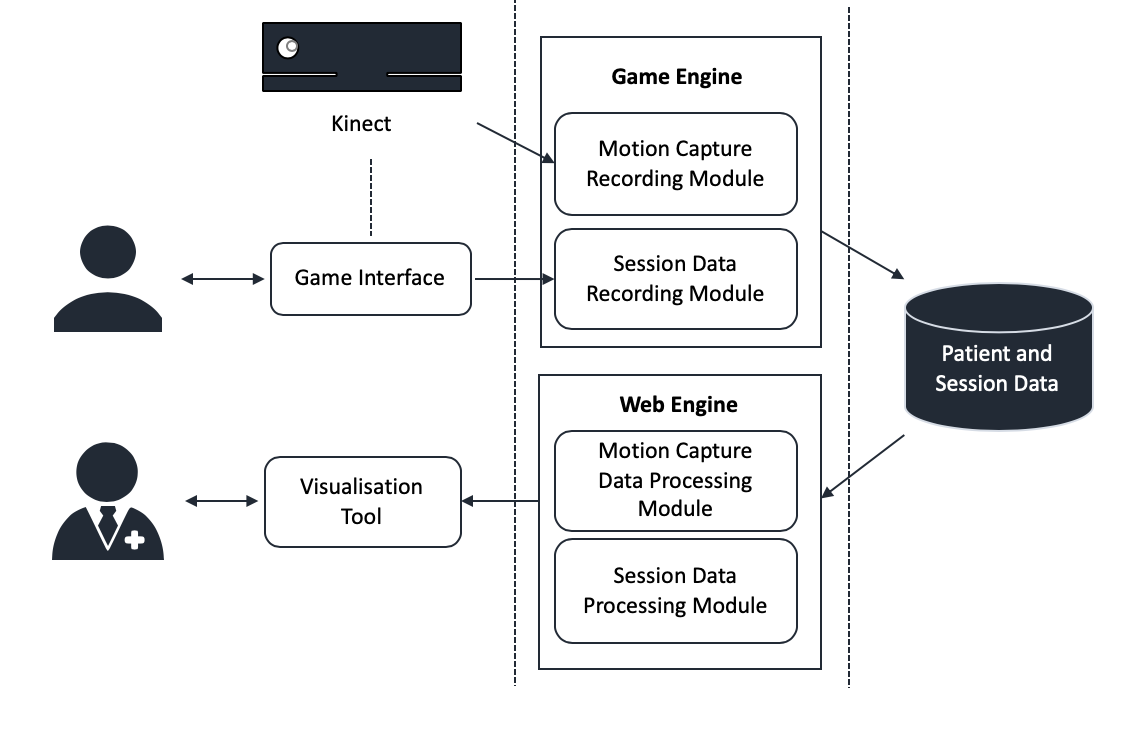}
    \caption{Overview of Tele-rehabilitation System}
    \label{fig:sysoverview}
\end{figure}

This web tool makes use of data captured while users are playing a Serious Game called MoMu \cite{agres2017music}. The game engine records motion capture data from the Kinect, as well as session-related data such as game score, the audio track (if any) used for each exercise, and the movement limits. The motion capture data is stored in .trc format files, and stores the $x$, $y$ and $z$-coordinates of all 25 joints that the Kinect tracks. This data is then processed in the back-end of a web server in Python before being projected in the front-end in the form of tables and visualisations. The front-end was based upon but modified from a MIT-Licensed template obtained from   \httpAddr{creative-tim.com/product/paper-dashboard}. The graphs displayed in this tool are visualised using the open-source charting library, Chart.js, which can be obtained from \httpAddr{//chartjs.org}. 
Figure ~\ref{fig:sysoverview} briefly displays the flow of the entire system. 

This work focuses on the interface of this new web tool, the computation of the kinematic measures, and the visualisation of these measures.

The objective of this tool is to allow healthcare professionals to remotely monitor a patient's progress in a rehabilitation regimen via tele-rehab. To fulfil this goal, the finer requirements of the system are to provide 1) insight on how frequently a patient is engaging with the system, 2) comparisons of movement quality over time, 3) insight on movement quality across trials within a session, and 4) kinematic information within a particular trial.  

To facilitate the above requirements, our tool provides information on a patient's performance at three levels: patient-level, session-level, and trial-level.
The patient-level overview of the patient's interaction with the system records information on the number of sessions the patient has completed, the total amount of time the he has engaged in exercises using the system, and the current average score. Details of the sessions such as time and date, along with kinematic measures such as the mean velocity, average smoothness score, and average auto-correlation value of all trials in the session are presented in the form of a table. A chart view, which visualises the patient's engagement with the system over the weeks/months is accessible through the 'Chart View' button.

From this high-level view of the patient, the web tool then allows for healthcare professionals to scroll down to the session-level, either by clicking on an element in the chart or on a row in the table. At the session level, healthcare professionals may see the trends in auto-correlation and smoothness scores across trials in the session. These metrics were chosen because they are arguably the most effective in conveying the level of consistency and the quality of movement of the patient. This overview is useful to examine, for example, when the participant's performance peaks during the session, or at which point the patient's performance starts declining, if at all. With this feedback, healthcare professionals may be better able to customise exercise sequences for their patients.  

Finally, the trial-level, the most detailed level of the tool, provides visualisations of several kinematic measures for individual trials. At this level, healthcare professionals have a view of how a patient's performance varies over the course of a single trial.

\begin{figure*}[htp]
        \centering
        \includegraphics[width=\textwidth]{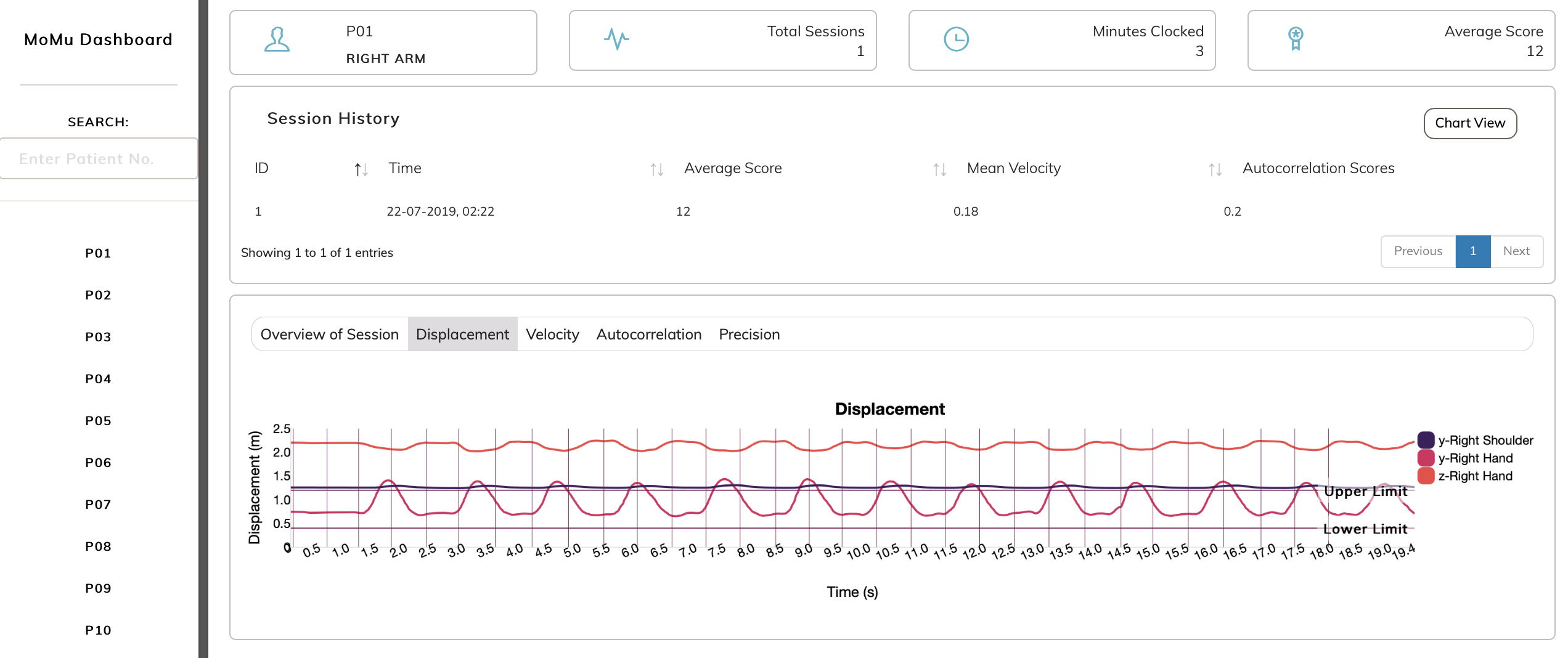}
        %\label{fig:autocorr1}

    \caption{Screenshot of Web Tool}
    \label{fig:sessionlog}
\end{figure*}

Figure ~\ref{fig:sessionlog} shows a screenshot of the dashboard displaying the highest level of detail, the patient profile overview, down to the lowest, the visualisation of a kinematic measure for a specific trial. A sidebar allows healthcare professionals to search for a specific patient whose profile they would like to view.
%-------------------------------------------------------------------------
\section{METHOD}

\subsection{Participants}
Seventeen adults participated in the study with an average age of 68 years (SD = 5.1 yrs). The musical experience of participants ranged from none up to 60 years of musical experience. The participants involved in this study were physically active with regularly scheduled physical activities.

\subsection{Procedure}

Participants were given a verbal explanation of the study, as well as written information sheet explaining the study and the task. After informed consent was obtained, the participant stood approximately 2m in front of the Kinect camera, facing a large computer monitor to interact with the system. The start and stop heights for the abbduction-adduction movements were marked by tennis balls attached to a stand next to the participants. An avatar was displayed on the screen which mirrored the participant's movements. 

During the study, participants performed a shoulder adduction/abduction exercise for 15 to 20 seconds per trial. Abduction is the movement of a body part (in this case, the right arm) away from the midline of the body, while adduction is the movement of the part towards the body.
In these exercises, participants began with their right arm extended straight out to the side from their shoulder and then moved their arm towards their body, before bringing their arm back to shoulder height again. This movement pattern was repeated until the end of the trial. The height at which a participant begins the exercise with their hand extended at the shoulder is marked as the 'Upper Limit', and the lowest position of the hand during the movement is marked as the 'Lower Limit'. These limits are used in the visualisations. In each trial, the first four visually cued beats of the music allow users to get accustomed to the tempo of the track. Therefore, the computation of auto-correlation and smoothness scores are based on data from the fifth beat till the end. Each participant went through 6 such trials, all within one session.

\subsection{Materials}
Royalty-free music tracks obtained from Audioblocks were used for the exercises. Resources were obtained from: \httpAddr{//www.audioblocks.com }. The tracks used were \textit{Barnyard Bash}, \textit{Makes Me Happy}, and \textit{Family Outings}.
Two versions of each track were used, one with a tempo of 80bpm and another with a tempo of 100bpm. Two silent tracks were also added to the list. The two silent tracks had visuals aligned to the 80bpm and 100bpm versions of the song `Family Outings'. 

Finally, the visualisations in this paper are based on the movement of two participants to the 100bpm version of the track entitled `Makes Me Happy'.
%-------------------------------------------------------------------------
\section{VISUALISATIONS AND ANALYSES }

This tool visualises various measures that we believe provide insight into the user's engagement with the system and the progress of a patient's quality of movement over sessions.
For instance, kinematic visualisations of the trajectory of the user's movements are provided (displacement). Other metrics involve velocity, auto-correlation and the amplitude spectrum. A session summary is also provided summarising the auto-correlation and smoothness of the user's movement for a particular session.

\subsection{Session Overview}
A general summary of each session is provided in the form of a line chart, reflecting how the smoothness scores and auto-correlation values change over trials within the session. The auto-correlation value for each trial is obtained by computing the average of the absolute values of the auto-correlation vector. Selecting any element on the chart leads the user to the kinematic visualisations of that particular trial.

\subsection{Displacement}
The displacement chart visualises a trajectory of the user's right hand during the activity. Again, the user's task is to synchronise their adduction/abduction movements (i.e., hit the upper and lower limits) with the beats of the music. 

A common compensatory movement during this type of exercise is the elevation of the shoulder, and leaning to the contralateral side. Therefore, as seen in Figure \ref{fig:sessionlog}, the vertical displacement of the right shoulder (marked as y-Right Shoulder) is visualised to allow one to asses whether compensatory movements were made by the patient. The z-dimension of the patient's right hand is also displayed to ensure that the patient is executing the exercise correctly. Ideally while performing this exercise, there should be limited y-dimensional movement in the shoulders as well as z-dimensional movement of the hand. Horizontal lines indicate the Upper and Lower limits participants are required to hit during game play, and vertical lines indicate the timings of beats of the audio track.
The coordinate vectors for each of the 25 markers and dimensions (x, y and z) obtained from the Kinect sensor are smoothed with SciPy's implementation of the Savitzky-Golay \cite{schafer2011savitzky} filter with window size 5 and polynomial order 0 to produce the displacement vectors.

\subsection{Velocity}
Several studies have used mean velocities as a means of quantifying motor recovery in stroke patients. \cite{mazzoleni2013robot,piron2004,malcolm2009}. \cite{nordin2014assessment} found that peak and mean velocity are indicative of ease of movement in stroke patients. \cite{thaut2002kinematic} has also suggested that smoother velocity profiles are indicative of rhythmic entrainment and improved movement quality in stroke patients. The velocity plot displays the highest velocities the user can attain, and where during the movement interval the peak occurs. \cite{zollo2011quantitative} found that as the quality of movement improves, the velocity curve moves towards the form of a symmetric bell curve. To obtain the velocity, the gradients of the displacement vector are obtained using NumPy's gradient function before multiplying the gradient vector by the frame rate. 

\vspace{-.2cm}
\subsection{Movement Smoothness}
Movement smoothness is another metric that is used to determine the quality of movement in rehabilitation. For this sytem, we chose the spectral arc length (SPARC) \cite{engdahl2019reliability, husaain2016} to quantify movement smoothness. The SPARC is computed as follows:
\vspace{-.2cm}
\begin{multline}
    \text{SPARC} = -\int_{0}^{\omega_{c}}\!\left[\!\!\left(\!\frac{1}{\omega_{c}}\!\right)^{2} \,+\, \left(\!\frac{d\hat{V}(\omega)}{d\omega}\!\right)^{2}\!\right]^{\frac{1}{2}}\!d\omega;\\
    \hat{V}(\omega) = \frac{V(\omega)}{V(0)},\\
    \omega_{c} \triangleq \min\left\{\omega_{c}^{max}, \min\left\{\omega\,\,\,,\hat{V}(r)<\overline{V} \,\,\, \forall \,\,\, r >\omega \right\} \right\}
\end{multline}

where $V(\omega)$ is the fourier magnitude spectrum of the velocity signal $V(t)$, and $\hat{V}(\omega)$ is the normalised magnitude spectrum \cite{balasubramanian2015analysis}.

The smoothness of the movement is computed by first breaking the movement down to cycles of one adduction and one abduction movement each. This is accomplished by identifying the lowest and highest values in each abduction-adduction cycle. 
The peaks are required to have a vertical height of at least the mean of the signal, and a horizontal distance of the time period of two beats. The lowest value between the peaks is then computed to mark the end of the adduction movement. 
The SPARC value is computed for each of these sub-movements, and the average of these SPARC values is used to determine the smoothness value for each trial.

\subsection{Auto-correlation}

We use the auto-correlation function to give us a visualisation of the periodicity of the user's movement. This measure gives insight on the consistency of a user's movement throughout the trial. In this case, auto-correlation was computed using NumPy's correlate function, which uses the definition as follows, where $k$ refers to the lag, and $a$ and $v$ are two different vectors. Because we are computing the correlation of a vector with itself, $a$ and $v$ were set to be the same.
The auto-correlation values are then normalised to allow for comparison between sessions. 
\begin{equation}
Correlation = \sum_n \left(a\left[n+k\right]*v\left[n\right]\right)
\end{equation}
\vspace{-.2cm}
\begin{figure}[ht]
    \centering
    \includegraphics[width=\columnwidth]{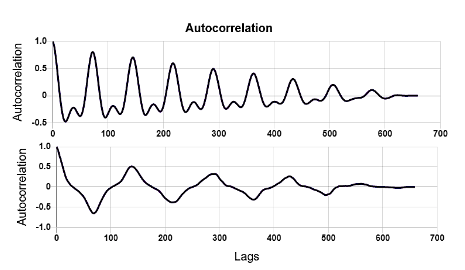}
    \caption{Auto-correlation Charts}
    \label{fig:autocorrchart}
\end{figure}

The charts in Figure \ref{fig:autocorrchart} display the normalised auto-correlation values of two participants. Here, in the first graph, the peaks  are higher and the auto-correlation decays more slowly than in the second graph. We can infer from these results that there is a higher level of periodicity in the first case than the second. 

\subsection{Amplitude Spectrum}

The amplitude spectrum is obtained by computing the magnitude of the Discrete Fourier Transform (DFT) of the displacement vector. The DFT is computed using Numpy's implementation which is as follows:
\vspace{-.2cm}
\begin{equation}
    A_k = \sum_{m=0}^{n-1} a_m e^{-2\pi i ^{\frac{mk}{n}}}
\end{equation}

The function produces a complex array $A$ where the first element is the sum of the signal, $A[1:n/2]$ contains the positive-frequency terms, and $A[n/2+1:]$ produces the negative-frequency terms. For the visualisation of the amplitude spectrum, because only one side of the Fourier transform is being presented in the graph, the amplitude is multiplied by 2. 

% The intention of this graph is to not only show periodicity, but also to display whether users are moving periodically at the desired frequency. As seen in Figure \ref{fig:ampspecchart}, the dominant peaks in both graphs show that there is some periodicity in the movement of both users. We can note that in both cases, the users are moving at a frequency slower than the ideal, although the first user is closer to the ideal frequency than the second. The sharper and higher dominant peak in the first graph indicates that the user's movement is more precise and consistent. 

%-------------------------------------------------------------------------
\subsection{Conclusion}

We believe that the visualisations and kinematic measures implemented here in the form of a web platform are useful in conveying differences and progress in quality of movement in patients performing motor rehabilitation exercises with auditory cueing. Besides using several dimensions to explore a patient's quality of movement, this tool also accommodates the use of music in the rehabilitation process and revolves around using rhythmic entrainment for motor rehabilitation, which, to the best of our knowledge, are features that are not present in any existing visualisation tools for motor tele-rehabilitation. 

The graphs of the various metrics display the extent to which the user's movements are optimised, whether they are able to reach the designated limits in time, and the regularity of their movements. Summary metrics such as mean velocity, smoothness, and auto-correlation scores quantify the consistency and quality of the user's movements over sessions. Future work aims to validate these metrics and visualisations with patients, and implement new visualisation features which would be useful to healthcare professionals. We also plan to incorporate analysis of other forms of rehabilitation movement and exercises into the platform.

%-------------------------------------------------------------------------

% bibtex
\bibliographystyle{eg-alpha-doi}
\bibliography{viz_paper}

% biblatex with biber
% \printbibliography                

%-------------------------------------------------------------------------
\newpage

\end{document}